\newcommand{\be}{\begin{equation}}
\newcommand{\ee}{\end{equation}}
\newcommand{\ba}{\begin{eqnarray}}
\newcommand{\ea}{\end{eqnarray}}
\newcommand{\n}{\nonumber \\}
\newcommand{\eq}[1]{(\ref{#1})}

\newcommand{\tGamma}{{\widetilde \Gamma}}

\documentclass[12pt]{article}
\begin{document}

\begin{titlepage}
\nopagebreak
\begin{flushright}
May 1999\hfill
UT-838\\
hep-th/9905044
\end{flushright}

\vfill
\begin{center}
{\LARGE Fate of unoriented bosonic string}\\
{\LARGE after tachyon condensation}
\vskip 20mm

{\large Yutaka MATSUO}
\vskip 5mm
{\sl APCTP}\\
{\sl 207-43 Cheongryangri-dong, Dongdamun-gu}\\
{\sl Seoul 130-012, Korea}\\
\vskip 2mm
{\em and}
\vskip 2mm
{\sl Department of  Physics, University of Tokyo}\\
{\sl Hongo 7-3-1, Bunkyo-ku, Tokyo 113-0033}\\
{\sl Japan}
\end{center}
\vfill

\begin{abstract}
By a toroidal compactification with 
a vortex like configuration of tachyon fields,
the unoriented bosonic string in 26 dimensions becomes
equivalent to 10 dimensional string theory with 
gauge group $SO(32)\times G$ where $G$ has rank 16.
The reduction from $SO(8192)$ 
to $SO(32)$ is induced by the action of non-abelian
Wilson lines which twists the tachyon.
The additional enhanced gauge symmetry $G$ 
appears from closed string sector by the vertex operator construction.
We also examine the consistency of the tachyon condensation.
\end{abstract}

\vfill
\end{titlepage}

\section{Introduction}

Some years ago, it was 
found  \cite{r:Weinberg} \cite{r:DG} \cite{r:MS}
that the unoriented
bosonic string theory
with gauge group $SO(2^{13})=SO(8192)$
meets tadpole cancellation condition.
Although it is interesting to observe that 
a consistent string theory exists with such a simple field content,
it is far from realistic because of the existence of tachyon and
the large gauge group.

Recent development of string theory show that a paradigm that
every tachyon free, modular invariant string theories
(type IIA/B, type I, and heterotic string)
should be regarded as different
limits of moduli of one unifying model, say M theory.
At this moment, however, it is still difficult to
understand the status of the  modular invariant but tachyonic
theories such as bosonic, type O, OA/B, and some heterotic
models. The key issue is how to make physical interpretation
of tachyons.

Starting from the study of stable non-BPS sates 
\cite{r:Sen1}-\cite{r:Sen6}, Sen began to clarify
the meaning of tachyon. In parallel
D-brane - anti-D-brane system, tachyon plays the r\^ole of
the Higgs fields  in the system with spontaneous symmetry breaking. 
By tachyon condensation with topologically 
nontrivial winding,  D-$(p-2)$-brane appears at the 
core of vorticity after ``pair annihilation'' of
D- and anti D-branes. A geometrical basis
of such a mechanism was explored by Witten \cite{r:Witten}
in  the language of K-theory.

The purpose of this letter is to analyze a possibility
that such a mechanism may be applied to bosonic $SO(8192)$ theory.
The idea is very simple.  Originally, the theory has $2^{13}=8192$
25-branes.  After the tachyon condensation, the number of branes 
should be reduced by halves and 
dimensions of branes will be decreased by two. 
If we repeat this step 8 times, we get reduction of
$2^{13}$ branes to $2^5=32$ branes. 
If we choose the momentum lattice of toroidal
compactification,  an extra gauge symmetry $G$ of rank 16
may appear from the closed string sector.  
If we take $G=SO(32)$ the system has the same gauge group
as that of the type 0 theory \cite{r:BG} 
and toroidal compactifications of oriented bosonic closed string.
In this way, one may include $SO(8192)$ theory into
a family where a reasonable treatment of remaining
tachyon is being developed recently from AdS/CFT 
correspondence \cite{r:KT}.

In \cite{r:Sen7} Sen has created a method
to treat tachyon condensation as a deformation
of boundary conformal field theory (BCFT).
By following him, we will use 
three steps to explicitly describe tachyon condensation.
\begin{enumerate}
\item Find a insertion of Wilson line which
will make tachyon mode anti-periodic  in the
compactified directions. It creates a kink in that direction.
\item Find a radius where tachyon vertex has dimension one
where it becomes marginal operator.
\item Realize tachyon condensation as the deformation of BCFT.
\end{enumerate}
Sen \cite{r:Sen7} has developed BCFT of tachyon condensation
where the dimension of the brane will be 
reduced by one by merging two D-branes.
Straightforward repetition of his method will reduce the number of
branes by a factor of $2^{16}$ while reducing the dimension 
of the branes by $16$. Since we have only $2^{13}$ branes,
it should be modified for our purpose.  Namely, we need
to study the process of pair annihilation of two D-$(p+2)$ branes
to create one D-$p$-brane just as in the superstring cases
\cite{r:Sen4}\cite{r:Witten}\cite{r:Sen5}.

\section{Generalized kink configuration and nonabelian Wilson lines}

For simplicity, we start from the reduction of two D-$(p+2)$-branes.
Chan-Paton (CP) group is $U(2)$ and we have four CP 
sectors labeled by $I$ and $\sigma_i$ ($i=1,2,3$).
We write two directions where reduction occurs as $x^1,x^2$.
At the core of the vorticity, tachyon should behave as\footnote{
We used the analogy with the superstring case \cite{r:Witten}.
In bosonic situation, this does not necessarily comes
from topological requirement since both tachyons and gauge
bosons has $U(2)$ CP factors.  Rather we use it simply
as a  generalization of the kink configuration.}
\be\label{e:core}
T(x)= x^1 \sigma_1 + x^2 \sigma_2+O(|x|^2).
\ee
Let us assume that $x^1$ and $x^2$ directions are
toroidary compactified in a radius $R$ and try
to achieve the first step of Sen's scenario.
It will be natural to assume from \eq{e:core}
that tachyon field $T(x)$ is anti-periodic
in $x^i$ direction when its CP factor is $\sigma_i$
(resp. $i=1,2$). We need to find Wilson
lines which generate such a change of 
boundary condition.
We denote $X^i(z)$ the string coordinates 
along the directions $x^i$. Consider the insertion of
Wilson line,
\be
\exp\left( i \frac{\alpha}{4} \sigma_2
 \oint \partial_z X^1(z)dz \right),
\ee
at the boundary of the world sheet.
Although the states with CP factors $I$ and $\sigma_2$
are unaffected, those with CP factors $\sigma_1$ and $\sigma_3$
get some shift of momentum $p_1\rightarrow p_1\pm\frac{\alpha}{2}$,
because of the commutation relation,
\be
[\sigma_2, \sigma_1\mp i \sigma_3]= 
\pm 2(\sigma_1\mp i \sigma_3).
\ee

More explicitly, if we write the zero-mode wave function 
of the open string sector as 
\be
\Psi(x^1,x^2) = \sum_{i=0}^3 \psi_i(x^1, x^2) \sigma_i,
\quad \sigma_0=1,
\ee
the boundary condition is modified to,
\be
\Psi(x^1+2\pi R,x^2) = h^1_\alpha \Psi(x^1,x^2) 
(h^1_\alpha)^{(-1)},\qquad
h^1_\alpha = \exp(i \frac{\pi}{2} R\alpha \sigma_2).
\ee

Momentum along $x^1$ was originally quantized in 
the unit $1/R$. Therefore if we choose $\alpha=1/R$
we obtain desired anti-periodicity
for the mode with CP factor $\sigma^1$. The factor 
in the boundary condition becomes $h^1=\sigma_2$.

To get correct anti-periodicity along $x^2$ direction,
we need to include the second Wilson line,
\be
\exp\left( i \frac{\alpha}{4} \sigma_1
 \oint \partial_z X^2(z) dz \right).
\ee
This time those modes with CP factor $\sigma_2$ and $\sigma_3$
get similar shift of their momentum. 
\be
\Psi(x^1,x^2+2\pi R) = h^2_\alpha \Psi(x^1,x^2) 
(h^2_\alpha)^{(-1)},\qquad
h^2_\alpha = \exp(i \frac{\pi}{2} R\alpha \sigma_1).
\ee

We choose $\alpha=1/R$ again and the deformation factor becomes
$h^2=\sigma_1$. The deformations factor for each Wilson line
become non-commutative $h^1 h^2\neq h^2 h^1$.  However, the action 
on the wave function becomes commutative and there is no
ambiguity in the ordering. This happens because of the
specific choice of $\alpha$ we made.

At this point it is worth while to discuss the topological
aspects of our gauge configuration. If we convert the Wilson 
lines to the gauge field, it gives constant gauge fields,
\be
A_1(x)=\frac{\alpha}{4}\sigma_2 \qquad
A_2(x)=\frac{\alpha'}{4}\sigma_1
\ee
Unlike the usual situation in the Wilson line,
it gives the nonvanishing curvature
in the gauge background,
\be
F_{12} \sim \sigma_3.
\ee
In this sense, the introduction of the second Wilson line
is not achieved by the continuous deformation but should be
regarded as the discrete transformation.

In the superstring case \cite{r:Sen6}\cite{r:Witten}, 
the topologically distinct sectors are labeled by
$\pi_1(U(1))={\bf Z}$. It gives the D-$(p-2)$-brane charge at the core
of vorticity. On the other hand in the bosonic situation,
it seems rather doubtful that we may have topologically 
distinct sectors since $\pi_1(SU(2))=0$.  However we have
to note that the open string wave function transforms in the
adjoint representation of $SU(2)$ and is topologically
parameterized  by $SO(3)$. Since $\pi_1(SO(3))=Z_2$, we have
a topologically nontrivial sector. The gauge configuration
that we have constructed  exactly belongs to this sector.
This statement can be verified by observing that the monodromy
matrices did not commute with each other $h^1h^2=-h^2h^1$.
It gives the distinct elements in $SU(2)$ but results in the consistent
boundary condition for the $\Psi$.
Since our configuration is topologically nontrivial, it
should be regarded as the stable configuration of the
open string theory.

To summarize the situation of the boundary condition
for the wave function, if we write $(\pm, \pm)$
as the sign of (anti-)periodicity in $x^1$ and $x^2$ direction
respectively, four CP sectors are split into,
\be\label{e:periodicity}
I: (++) \quad \sigma_1: (-+) \quad \sigma_2: (+-) \quad \sigma_3: (--)
\ee

At this point, if we combine
all the modes from each CP factor, the momentum distribution 
is identical to those of rectangular lattice with radius $2R$
as far as open string sector is concerned .
If we take T-duality transformation in $x^1$ and $x^2$ directions,
one may alternatively view it as winding mode of open string 
originating from one D-$p$-brane for radius $1/2R$.
In this way, we obtained three equivalent descriptions 
for the same system.

\begin{enumerate}
\item  two D-$(p+2)$-branes with Wilson lines
	compactified with the radius $R$
\item  one D-$(p+2)$-brane with the radius $2R$
\item  one D-$p$-brane with the radius $1/2R$
\end{enumerate} 

Such an equivalence, however, may not be straightforwardly
interpreted as the usual tachyon condensation scenario
if we keep the closed string sector in our mind. Namely, the inclusion
of Wilson lines do not change the closed string spectrum at all.
Therefore, closed strings still  live in the space with the 
radius $R$ or its dual $1/R$. It does not fit with above 
{\em brane number reduction} scenario. To realize a real reduction, 
we need to keep the original radius $R$ invariant. Such a mechanism
occurs only when we introduce the deformation 
by tachyon fields. We will discuss this issue in section 4.

There is however a positive lesson from the above
observation. We may claim that the original 
$U(2)$ symmetry is broken to
$U(1)$ contrary to the usual situation where unbroken symmetry
is $U(1)\times U(1)$. The origin of such a phenomena
is that we introduced two Wilson lines in mutually
non-commuting CP directions $\sigma_1$ and $\sigma_2$.
For the reduction from $O(8192)$ to $O(32)$, this claim 
is already enough and we do not need full
tachyon condensation scenario.

While the gauge symmetry from open string sector
is reduced, we may choose the radius of the lattice
to realize a new gauge symmetry from closed string sector.  
By taking $R=1$, we have two pairs of $SU(2)$ current algebras,
\be
J^{[I]}_\pm=e^{\pm i X^{[I]}},\quad 
J^{[I]}_3=\partial X^{[I]},\qquad I=1,2.
\ee
This $SU(2)\times SU(2)$ becomes the gauge symmetry
of the space-time by using standard vertex operator 
construction.
\footnote{Gauge group is $SU(2)\times SU(2)$
instead of $SU(2)^{\otimes 2}_L\times SU(2)^{\otimes 2}_R$ 
since we consider unoriented strings and half of gauge
bosons are projected out.} 

\section{Application to $SO(8192)$ theory and reduction in one step}

Let us apply this idea to $SO(8192)$ model.
To make the discussion clearer, we compactify one extra dimension,
say $X^9$ and apply T-duality transformation along that direction.
If we introduced an appropriate Wilson line, those 8192
D-branes will be split into $4096=2^{12}$ D-branes located
at the same point, an orientifold and mirror images of D-branes.
The unbroken symmetry becomes $U(2^{12})$.
We may introduce non-abelian Wilson lines to reduce the
symmetry to $U(2^{11})$.
While repeating above procedure for 8 times, the number of branes
become $2^{12-8}=2^4=16$. Together with orientifold plane, 
the surviving D-branes define $SO(32)$ theory.
The extra gauge symmetry from the closed string sector
becomes $SU(2)^{\otimes 16}$ if $R=1$.

To compare this ten dimensional model with type 0 
string theory, it is necessary to deform 
momentum lattice $SU(2)^{\otimes 16}$.

If we deform the lattice, it is not possible to use
step by step method. Rather we need to develop 
the reduction of space-time in one step \cite{r:Witten}.
If we write $x^I$ ($I=1\cdots 16$) our compact directions
and $X^I$ as the string variable associated with it.
First let us consider rectangular 16 dimensional lattice
again. We prepare an explicit representation for gamma matrices
in 16 dimensions (256 dimensional representation) as,
\ba
\Gamma_1  =  \sigma_1\otimes 1\otimes \cdots \otimes 1&\qquad&
\Gamma_2  =  \sigma_2\otimes 1\otimes \cdots \otimes 1\n
\Gamma_3  =  \sigma_3\otimes \sigma_1\otimes \cdots \otimes 1&\qquad&
\Gamma_4  =  \sigma_3\otimes \sigma_2\otimes \cdots \otimes 1\n
\cdots &&\n
\Gamma_{15}  =  \sigma_3 \otimes \sigma_3\otimes \cdots \otimes \sigma_1
&\qquad &
\Gamma_{16}  =  \sigma_3 \otimes \sigma_3\otimes \cdots \otimes \sigma_2.
\ea
Together with these sets,  we also need another representation,
\ba
\tGamma_1  =  \sigma_2\otimes \sigma_3\otimes \cdots \otimes \sigma_3 &\qquad&
\tGamma_2  =  \sigma_1\otimes \sigma_3\otimes \cdots \otimes \sigma_3\n
\tGamma_3  =  1\otimes \sigma_2\otimes \cdots \otimes \sigma_3 &\qquad&
\tGamma_4  =  1\otimes \sigma_1\otimes \cdots \otimes \sigma_3\n
\cdots &&\n
\tGamma_{15}  =  1 \otimes 1\otimes \cdots \otimes \sigma_2&\qquad&
\tGamma_{16}  =  1 \otimes 1\otimes \cdots \otimes \sigma_1.
\ea
These matrices are chosen to satisfy,
\be
[\Gamma_I, \Gamma_J]_+=[\tGamma_I, \tGamma_J]_+=2 \delta_{IJ}, 
\quad
[\Gamma_I,\tGamma_I]_+=0, \quad
[\Gamma_I,\tGamma_J]_-=0 \quad (I\neq J).
\ee
The inclusion of Wilson line,
\be\label{e:Wilson}
\prod_{I=1}^{16} \exp\left(
	\frac{i}{4R} \tGamma_I \oint \partial_z X^I dz\right),
\ee
will modify the periodicity of tachyon in CP factor $\Gamma_I$ to
$(+,\cdots,+,-,+,\cdots,+)$ where minus sign occurs only at $I$ th position.
In other word, we may have condensation of tachyon in the 
following form \cite{r:Witten} at $x^I=0$,
\be
T(x) = \sum_{I=1}^{16} x^I \Gamma_I +O(|x|^2).
\ee
Inclusion of Wilson line \eq{e:Wilson} will invoke
$2^{16}$ sectors of CP factor split into $2^{16}$ points
in the half rectangular lattice.

It is now straightforward to generalize the argument
to the skew lattice case.  To get gauge bosons from
the closed string sector, we need to restrict it
to a self-dual lattice. We denote $\vec E_I$ is
the basis of  lattice and $\vec F_J$ as the dual
basis $\vec{F}_I \cdot \vec{E}_J = \delta_{IJ}$.
The tachyon mode at the origin should behave as,
\be
T(x) = \sum_{I} \Gamma_I \sin(\vec E_I\cdot \vec x)
\sim \sum_{I} \Gamma_I (\vec{E}_I\cdot \vec x)\qquad
(x\sim 0).
\ee
Wilson line that invokes such a tachyon mode is
\be
\prod_{I=1}^{16} \exp\left(
	\frac{i}{2} \tGamma_I  \oint \vec{F_I}\cdot 
\partial_z \vec X dz\right).
\ee
To obtain $SO(32)$, we take the original lattice as
the $SO(32)$ lattice. 

\section{Tachyon condensation scenario}
We come back to the consideration of the reduction
by two dimensions.
So far, there was no restriction on $R$ from
the open string sector.  To examine the scenario
of tachyon condensation, we need to impose the
tachyon vertex to have dimension one.

Tachyon fields in twisted direction have mode expansion,
\be
T(x)=\sum_{m\in Z} T_{m+1/2} e^{i(m+1/2)x}.
\ee
The mass squared of $T_{\pm 1/2}$ mode is enhanced to,
\be
m^2 = \frac{1}{4R^2} -1.
\ee
At $R=1/2$ they become massless and can be used 
to deform the boundary conformal field theory (BCFT)
\cite{r:Sen7}. 

The remaining analysis of tachyon condensation is almost
parallel to \cite{r:Sen7} and we will follow it step by step.
In this radius, the {\em anti-periodic} mode $e^{\pm i 2(X/2)}$
defines $SU(2)$ current algebra. 
The vertex operator for the tachyon condensation \eq{e:core} 
is exactly this non-diagonal $SU(2)$ current and 
we need to make nontrivial change of variable.
We introduce the translation operators,
\be
h_I: X^I \rightarrow X^I+\pi, \quad (I=1,2).
\ee
The eigenvalue of each CP sector under $h_I$ is given
in \eq{e:periodicity}. Together with $h$, we introduce
another $Z_2$ operator $g$ as,
\be
g_I: X^I \rightarrow -X^I.
\ee
From massless tachyon mode, we can construct
$SU(2)_L^{\otimes 2}\times SU(2)_R^{\otimes 2}$ 
current algebra,
\ba
e^{2i X_L^I}&=&\partial Y_L + i \partial Z_L\n
e^{2i Y_L^I}&=&\partial X_L - i \partial Z_L\n
e^{2i Z_L^I}&=&\partial X_L + i \partial Y_L,
\ea
for the left movers and the similar definition for the right movers.
We define $h$ and $g$ operator similarly for $Y$ and $Z$.
There are some relations between them,
\be
h_X=g_Y=g_Z,\qquad g_X=h_Y=h_Z g_Z.
\ee
The tachyon condensation is triggered by the 
inclusion of {\em non-Abelian} Wilson line again,
\be
\exp\left( i \frac{\alpha}{4} \oint dz
\partial Y_1(z)\sigma_1 
\right) 
\exp\left( i \frac{\alpha}{4} \oint dz
\partial Y_2(z) \sigma_2
\right). 
\ee
This time, since we originally have both $\pm 1$ eigenvalues
for the operator $h_{Y^I}$, it is more natural to follow
the change of the boundary condition in the translations,
$X^I \rightarrow X^I + 4\pi R$. Wilson lines shifts 
$h_{Y^I}^2$ eigenvalues
by $\exp(\pm 2 \pi i\alpha)$.  Since we use the double period,
shift of $\alpha$ by $1$ will not 
change the spectrum at all. At $\alpha=1/2$ they 
induces deformation factors $\sigma_{1,2}$ again on 
the wave function. Returning back to shift in $2\pi R$,
the eigenvalues of $h_Y$ becomes $\pm i$.
We summarize the eigenvalues for $g_Y$ and $h_Y$ in the following
table.

\begin{center}
\begin{tabular}{|c|c|c|c|c|}\hline
& $h_{Y^1}=g_{X^1}$ & $g_{Y^1}=h_{X^1}$ & $h_{Y^2}=g_{Y^2}$ 
& $g_{Y^2}=h_{X^2}$ \\ \hline
$I$ & $\pm 1$ & $+$ & $\pm 1$ &$+$ \\
$\sigma_1$ & $\pm 1$ & $-$ & $\pm i$ &$+$ \\
$\sigma_2$ & $\pm i$ & $+$ & $\pm 1$ &$-$ \\
$\sigma_3$ & $\pm i$ & $-$ & $\pm i$ & $-$ \\ \hline
\end{tabular}
\end{center}

Having four values $\pm 1, \pm i$ for $h_Y$ indicates
that the momentum are actually quantized in different radius $R=2$.
Of course in making such an identification we need to 
combine the contributions from all the sectors.
Taking T-duality, we obtain a theory with one
D-$p$ brane with compactification radius $1/2$.

Since we have come back to the original lattice,
the resultant theory may consistently couple to
the closed string sector. Of course, since
we made non-trivial change of variables for
the open string sector, we need to carry out similar
task for closed string sector which seems to be
quite non-trivial.

More important problem is that 
if we study the table more carefully on $g_Y$
eigenvalues, we immediately notice that we 
missed actually $3/4$ of the states!  For example, 
for $h_Y$ eigenvalue $(\pm, \pm)$, $g_Y$ eigenvalues
are fixed as $(+,+)$ and eigenfunction should be parity
even for $Y$ variables. Of course, there are no such restriction
for the toroidal compactification in $R=1/2$.

This phenomena may look like an artifact of the complicated
way of introducing non-commutative Wilson lines.
However it can be alternatively explained by a simple argument.
Originally we have four degeneracy for each momentum
because of four CP factors. If we want to arrive at
the change of the radius $R=1/2 \rightarrow 2$ 
in two directions, the number
of momentum mode will be multiplied by $4^2=16$.
We can not fill all of those states but only $1/4$ of them.

It is obviously necessary to clarify why such a missing states 
could come out.  A simple calculation reveals that there
seems to be also a similar problem in a superstring scenario.

\section{Discussions}

Although we can not use the usual scenario of the tachyon 
condensation, we have shown that $SO(8192)$ theory 
compactified to ten dimensions may have gauge symmetry
$SO(32)\times SO(32)$.
Unfortunately, the field content of this theory
is rather different from those theories.
Especially at the bottom of the spectrum, we still have
$528$ (symmetric representation of $SO(32)$) tachyons.
Since the number of the tachyon
is essential to determine the vacuum structure,
it is hard to believe that these theories will have 
common fate.

We are tempted to believe that 
the toroidal compactification
to make tachyon massless is indispensable for the
stability of the system.
In analogy with standard Higgs mechanism, 
this may be called as ``spontaneous
compactification'' of the space-time.
Although we discussed mainly the compactification 
to ten dimensions, we believe that
unoriented bosonic string will unavoidably experience
further compactification until
it has only one tachyon from open string mode.  
This means that we will have a compactification
of 24 dimensions to leave only two uncompactified space-time.
At this stage, CP factor is just $O(2)=U(1)$
but enhanced gauge symmetry may have rank 24.
The final fate of $O(8192)$ theory should be 
clarified by string theory living in two dimensions.
We have one real tachyon in the open string sector.
If one may deform it in kink configuration again,
we would get a quantum mechanics!
However, the last tachyon is neutral in all CP factors,
we can not find appropriate Wilson line to deform it.
The correct treatment is still mysterious for us.

\vskip 10mm
\noindent{\bf Acknowledgements:\hskip 5mm}
The author would like to thank T. Takahashi
for attracting his interest to this subject and 
having intriguing discussions.
He is also obliged to Soo-Jong Rey and Sang-Jin Sin
for inviting him to APCTP and for their warm hospitality.


\end{document}